\documentclass[a4paper,11pt]{article}
\pdfoutput=1 

\usepackage{jcappub} 
                     
\usepackage{orcidlink}
\usepackage[T1]{fontenc} 
\def\be{\begin{equation}}
\def\ee{\end{equation}}
\def\bea{\begin{eqnarray}}
\def\eea{\end{eqnarray}}

\title{Classical and Loop Quantum Cosmology of Interacting Dark Energy: A Dynamical System Analysis with Superfluid Dark Matter and Dust Matter}

\author[a]{Koblandy Yerzhanov\,\orcidlink{0000-0003-0732-2080}}
\author[b,1]{Mohd Shahalam\,\orcidlink{0000-0003-3389-260X}\note{Corresponding author.}}
\author[c]{Gulnur Bauyrzhan\,\orcidlink{0000-0002-8410-6640}}
\author[d]{and Praveen Kumar Dhankar\,\orcidlink{0000-0002-8201-6019}}

\affiliation[a]{Department of General and Theoretical Physics, L. N. Gumilyov Eurasian National University, Astana 010008, Kazakhstan}
\affiliation[b]{Department of Physics, Integral University, Lucknow 226026, India}
\affiliation[c]{Department of Space and Technology, L. N. Gumilyov Eurasian National University, Astana 010008, Kazakhstan}
\affiliation[d]{Symbiosis Institute of Technology, Nagpur Campus, Symbiosis International (Deemed University),  Pune-440008, India
}

\emailAdd{mohdshahamu@gmail.com}
\emailAdd{yerzhanovkk@gmail.com}
\emailAdd{bauyrzhan.g.b@gmail.com}
\emailAdd{pkumar6743@gmail.com}

\abstract{We study the cosmological dynamics of interacting dark energy and dark matter in Classical Einstein Gravity and Loop Quantum Cosmology. Two dark matter scenarios are considered: superfluid dark matter described by a generalized cubic equation of state and the standard pressureless fluid. The dark energy component is modeled using both a generalized nonlinear equation of state and a constant equation of state. We examine two phenomenological interaction terms, $Q=\alpha\dot{\rho}_m$ and $Q=\beta\dot{\rho}_d$, which govern the energy transfer between the dark sectors. In classical gravity, the pressureless matter model exhibits stable late-time attractors, whereas the superfluid dark matter model admits only saddle and non-hyperbolic critical points. Extending the analysis to Loop Quantum Cosmology, quantum geometric corrections replace the Big Bang singularity with a nonsingular quantum bounce, and significantly modify the phase-space dynamics. As a result, the stable attractors of the classical pressureless matter model disappear, and all interacting models possess only saddle and non-hyperbolic critical points. These findings highlight the significant influence of both dark matter properties and quantum gravitational effects on the asymptotic evolution of interacting dark-sectors.}
\keywords{Classical gravity; Loop Quantum Cosmology; Interacting dark energy; Dynamical system analysis}
\arxivnumber{2607.01724}
\begin{document}
\maketitle
\flushbottom

\section{Introduction}
\label{sec:intro}
Observations of Type Ia supernovae, cosmic microwave background and large-scale structure, indicate that dark energy (DE) contributes approximately 70\% of the total energy density of the Universe, while dark matter (DM) accounts for about 25\%, leaving only a small fraction in the form of ordinary baryonic matter \cite{Riess:1998cb}. Although the $\Lambda$CDM successfully describes a broad range of cosmological observations, the fundamental nature of both DM and DE remains unknown. Moreover, the model faces cosmological coincidence problem and a number of small-scale discrepancies associated with the cold DM. Various theoretical approaches have been proposed to address these issues. Modified theories of gravity, particularly $f(R)$ gravity \cite{n1,n2,n3}, provide viable descriptions of both the inflationary era and the present accelerated expansion without invoking an explicit cosmological constant \cite{reviews1,reviews2,reviews3,reviews4,reviews5,reviews6,Nojiri:2003ft}. Alternatively, interacting dark-sectors assume a non-gravitational energy exchange between DM and DE, offering a possible explanation for the comparable energy densities of these components at the present epoch and leading to rich cosmological dynamics \cite{Gondolo:2002fh,Farrar:2003uw,Cai:2004dk,Bamba:2012cp,Guo:2004xx,Wang:2006qw,Bertolami:2007zm,He:2008tn,Valiviita:2008iv,Jackson:2009mz,Jamil:2009eb,He:2010im,Bolotin:2013jpa,Costa:2013sva,Boehmer:2008av,Li:2010ju,Yang:2017zjs}. In these models, the interaction is typically confined to the dark sector in order to avoid stringent observational constraints on couplings between DE and ordinary matter.

Most interacting dark-sectors describe DM as a pressureless fluid. However, an attractive alternative is provided by the superfluid dark matter (SFDM) \cite{Berezhiani:2015bqa,Berezhiani:2015pia,Hodson:2016rck,Berezhiani:2017tth}, in which DM undergoes Bose-Einstein condensation under suitable astrophysical conditions. This framework successfully reproduces the predictions of cold DM on cosmological scales while naturally giving rise to modified gravitational behavior on galactic scales, thereby addressing several long-standing small-scale problems of the $\Lambda$CDM, including the baryonic Tully-Fisher relation and the dynamics of dwarf galaxies \cite{Tulin:2017ara}. Consequently, SFDM provides a unified framework capable of explaining both cosmological and galactic observations. Motivated by these developments, we investigate cosmological models consisting of interacting dark energy (IDE) and DM within both classical Einstein gravity and Loop Quantum Cosmology (LQC). We consider two forms of non-gravitational interaction,
$Q=\alpha \dot{\rho}_m$ and $Q=\beta \dot{\rho}_d,$ where $\alpha$ and $\beta$ are dimensionless coupling constants \cite{Shahalam:2015sja, Shahalam:2017fqt}. For DM sector, we study both SFDM, described by a generalized polynomial equation of state \cite{Odintsov:2018uaw,Odintsov:2018awm}, and the standard pressureless matter (PM), allowing a direct comparison between the two scenarios.

Dimensionless dynamical variables are introduced to rewrite the cosmological equations as autonomous system \cite{Odintsov:2019ofr,Boehmer:2014vea,Bohmer:2010re,Sami:2012uh,a1,a2,a3,a4,a5,a6,a7,Goheer:2007wu,Leon:2014yua,Guo:2013swa,Leon:2010pu,deSouza:2007zpn,Giacomini:2017yuk,Kofinas:2014aka,Leon:2012mt,Gonzalez:2006cj,Alho:2016gzi,Muller:2014qja,Ivanov:2011vy,Boko:2016mwr,Odintsov:2017icc,Granda:2017dlx,Landim:2016gpz,Kleidis:2018cdx}. The stationary points are obtained by imposing vanishing derivatives with respect to the $e$-folding number, and their stability is determined through the eigenvalues of the Jacobian matrix.
In the classical framework, the phase-space analysis reveals a variety of stable and saddle  solutions. Depending on the interaction parameters, the system can evolve toward DE dominated, matter dominated, or scaling attractors. The presence of stable nodes demonstrates that interacting dark-sectors can naturally generate viable late-time cosmological states. To assess the impact of quantum gravitational effects, we extend the analysis to LQC \cite{LQC1,LQC3,LQC4,LQC5,Salo:2016dsr,Xiong:2007cn,Amoros:2014tha,Cai:2014zga,deHaro:2014kxa,Kleidis:2018plu,c1,c2,c3,c4,c5,Kleidis:2017ftt,Sami:2006wj}, where the modified Friedmann equation introduces non-perturbative quantum geometric corrections. These corrections substantially modify the global phase-space structure, resolve classical singularity through the quantum bounce, and alter the stability properties of the cosmological solutions. Although all critical points remain saddle. The paper is organized as follows. Section~\ref{sec:CC} presents the IDE and DM models in the classical gravity framework and analyzes four interaction scenarios involving SFDM and PM. Section~\ref{sec:LQC} extends the analysis to LQC and examines the corresponding quantum-corrected dynamical system. Section~\ref{sec:CI} is devoted to the cosmological implications for the models under consideration. Finally, the main results are summarized in Section \ref{sec:conc}.
\section{Classical Einstein gravity Framework}
\label{sec:CC}
We consider a cosmological model consists of IDE and SFDM in the framework of classical Einstein gravity. The DE and DM components are characterized by energy densities $\rho_d$ and $\rho_m$, respectively. For a spatially flat Friedmann-Lemaitre-Robertson-Walker (FLRW) spacetime, the first Friedmann equation is given by
\begin{equation}
H^2=\frac{\kappa^2}{3}\Big{(}\rho_d+\rho_m \Big{)},
\label{eq:H-CC}
\end{equation}
where $H=\dot a/a$ being the Hubble expansion rate, and $a(t)$ denotes the scale factor. The $\kappa^2=8\pi G$, with $G$ denoting Newton's gravitational constant. Assuming a non-gravitational interaction between the dark-sector components, the corresponding continuity equations take the form
\begin{align}
\label{eq:conser_1}
\dot{\rho}_m+3H(\rho_m+p_m)&=Q,\\
\dot{\rho}_d+3H(\rho_d+p_d)&=-Q,
\label{eq:conser_2}
\end{align}
where $p_m$ and $p_d$ represent the pressures of the DM and DE, respectively. The interaction term $Q$ governs the energy exchange between two sectors. A positive value of $Q$ corresponds to energy transfer from DE to DM, whereas $Q<0$ describes the opposite process. The Eqs. (\ref{eq:conser_1}) and (\ref{eq:conser_2}) indicate that $Q$ should be a function of $H$, $\rho_m$ and $\rho_d ~i.e.$ $Q(H, \rho_m, \rho_d)$. Following the phenomenological approach commonly adopted in the literature \cite{Nojiri:2005sx,CalderaCabral:2008bx,Pavon:2005yx,Quartin:2008px,Sadjadi:2006qp,Zimdahl:2005bk}, we choose two particular forms of interaction term as \cite{Shahalam:2015sja, Shahalam:2017fqt}
\begin{eqnarray}
\label{eq:Q1}
Q&=&\alpha \dot{\rho}_m,\\
Q&=&\beta \dot{\rho}_d,\\
\label{eq:Q2}
\end{eqnarray}
In these forms, $H$ is not directly involved, as it has the dimension of the inverse of time, and the latter is
already present in $\dot{\rho}_m$ and $\dot{\rho}_d$. Differentiating Eq.~(\ref{eq:H-CC}) and using the continuity equations yields the Raychaudhuri equation
\begin{equation}
\dot H=-\frac{\kappa^2}{2} \Big{(} \rho_m+\rho_d+p_m + p_d \Big{)},
\label{eq:H_d-cc}
\end{equation}
In the following subsections, we investigate the phase-space dynamics of four interacting models in the classical framework. The interaction between DM and DE is taken to be proportional to the time derivative of the DM density, $Q=\alpha\dot{\rho}_m$, and to that of the DE density, $Q=\beta\dot{\rho}_d$. For the DM sector, we consider both SFDM and pressureless matter (PM). For each model, we construct the corresponding autonomous system, determine its critical points, and analyze their stability through the eigenvalues of the Jacobian matrix. The cosmological implications of the stable and saddle solutions are discussed in the context of late-time evolution of the Universe.
\subsection{Model 1: $Q=\alpha \dot{\rho}_m$ with  Superfluid Dark Matter}
\label{sec:sub_A}
In this subsection, let us consider $Q=\alpha \dot{\rho}_m$, and work with  the SFDM scenario. Motivated by Ref.~\cite{Berezhiani:2015bqa}, we assume that the DM fluid obeys a cubic equation of state,
\begin{equation}
p_m=B \kappa^8\rho_m^3,
\label{eq:pm_GR}
\end{equation}
where $B$ is a dimensionless parameter. For the DE sector, we adopt the generalized equation of state \cite{sd}
\begin{equation}
p_d=-\rho_d-A\kappa^4\rho_d^2,
\label{eq:pd_GR}
\end{equation}
with $A$ being a dimensionless constant. To investigate the dynamical evolution of the cosmological system, we introduce the dimensionless variables
\begin{equation}
x=\frac{\kappa^2\rho_d}{3H^2},
\qquad
y=\frac{\kappa^2\rho_m}{3H^2},
\qquad
z=\kappa^2H^2.
\label{eq:variables}
\end{equation}
Using the number of $e$-folds $N=\ln a$ as the independent dynamical variable, the cosmological equations can be recast into the autonomous system
\begin{align}
\frac{dx}{dN} &= 9Ax^2z-\frac{\kappa^2 Q}{3 H^3}-2x \frac{\dot{H}}{H^2},\nonumber\\
\frac{dy}{dN} &= -3y-27By^3 z^2+\frac{\kappa^2 Q}{3 H^3}-2y \frac{\dot{H}}{H^2},\nonumber\\
\frac{dz}{dN} &= 2z \frac{\dot{H}}{H^2},
\label{eq:auto}
\end{align}
where,
\begin{eqnarray}
\label{eq:KQ1_SDF}
\frac{\kappa^2Q}{3H^3} &=& \frac{3\alpha}{\alpha-1} \Big{(}y+9 B y^3  z^2  \Big{)},\\
\frac{\dot{H}}{H^2} &=& -\frac{3}{2}\Big{(}y+9 B y^3  z^2 -3Ax^2 z \Big{)}
\label{eq:Hd1}
\end{eqnarray}
The effective equation of state parameter is defined as
\begin{equation}
w_{eff} = -1-\frac{2\dot{H}}{3H^2}
\label{eq:weff1}
\end{equation}
It is well known that generalized DE equation of state (\ref{eq:pd_GR}) may lead to finite-time cosmological singularity. Similar behavior has been reported in IDE model containing PM \cite{Odintsov:2018uaw}. The presence of a SFDM component modifies the dynamical structure of the phase space and may significantly affect the occurrence and nature of singularities.
The stationary (critical) points are obtained by setting the left-hand side of the autonomous equations (\ref{eq:auto}) equal to zero. The stability of each point is determined by analyzing the eigenvalues of the corresponding Jacobian matrix. The critical points of the system are listed below.
\begin{enumerate}
\item[\textbf{$P_1$:}] 
    \bea
   x = 0,  \qquad y = 0, \qquad z = 0,
    \eea
The eigenvalues of the Jacobian matrix evaluated at this point are
\bea
\eta_1 &=& 0, \nonumber \\
\eta_2 &=& 0, \nonumber \\
\eta_3 &=& \frac{3}{\alpha-1}
\label{eq:EVP1}
\eea
This point corresponds to a trivial state in which both the DE and DM densities vanish. Since two eigenvalues are zero, the point is non-hyperbolic and its stability cannot be determined from linear analysis alone.
\item[\textbf{$P_2$:}]
    \bea
    x=\frac{\alpha}{\alpha-1}, \qquad y =-\frac{\alpha}{\alpha-1} , \qquad  z = 0, 
    \eea
The corresponding eigenvalues are
    \bea
\eta_1 &=& \frac{3}{\alpha-1},\nonumber \\
\eta_2 &=& -\frac{3}{\alpha-1},\nonumber \\
\eta_3 &=& -\frac{3}{\alpha-1}
\label{eq:EVP2}
\eea
This solution represents a mixed dark-sector configuration. For $\alpha>1$, the eigenvalues possess opposite signs, indicating the presence of both stable and unstable directions. Hence, $P_2$ is a saddle point. Both points are summarized in Table \ref{tab:sub_A}.
\end{enumerate}
\begin{table}[tbp]
\centering
\begin{tabular}{|c|c|c|c|c|c|c|c|c|}
\hline
Point & \qquad $x$ & \qquad $y$ & \qquad $z$ & \qquad Stability & \qquad 
$w_{eff}$ \\
\hline
$P_1$ & \qquad 0 & \qquad 0 & \qquad 0 &  \qquad Undetermined  & \qquad
$-1$ \\
$P_2$ & \qquad $\frac{\alpha}{\alpha-1}$ & $-\frac{\alpha}{\alpha-1}$ & \qquad 0 & \qquad  Saddle  & \qquad 
$-\frac{2\alpha-1}{\alpha-1}$ \\
\hline
\end{tabular}
\caption{Critical points for Model 1, see subsection \ref{sec:sub_A}. The point $P_1$ is undetermined due to the non-hyperbolic nature in linear stability theory.}
\label{tab:sub_A}
\end{table}
\subsection{Model 2: $Q=\beta \dot{\rho}_d$ with  Superfluid Dark Matter}
\label{sec:sub_B}
We now consider the SFDM with a different interaction term, $Q=\beta \dot{\rho}_d$. In this case, the autonomous system (\ref{eq:auto}) and Eq. (\ref{eq:Hd1}) remain unchanged, whereas Eq. (\ref{eq:KQ1_SDF}) is modified due to the new coupling function, and can be expressed as
\begin{eqnarray}
\frac{\kappa^2Q}{3H^3} &=& \frac{9 A \beta z x^2}{1+\beta}
\label{eq:KQ2_SDF}
\end{eqnarray}
To determine the stationary points and examine their stability properties, we analyze Eq. (\ref{eq:auto}) together with Eqs. (\ref{eq:Hd1}) and (\ref{eq:KQ2_SDF}). The resulting critical points are as follows:
\begin{enumerate}
\item[\textbf{$Q_1$:}] 
    \bea
   x = 0,  \qquad y = 0, \qquad z = 0,
    \eea
The eigenvalues of the Jacobian matrix evaluated at this point are
\bea
\eta_1 &=& 0, \nonumber \\
\eta_2 &=& 0, \nonumber \\
\eta_3 &=& -3,
\label{eq:EVQ1}
\eea
Similar to $P_1$, this is a non-hyperbolic point corresponding to vanishing energy densities. Its stability remains inconclusive within linear stability theory.
\item[\textbf{$Q_2$:}]
    \bea
    x=0, \qquad y =1 , \qquad  z = 0, 
    \eea
The corresponding eigenvalues are
    \bea
\eta_1 &=& 3,\nonumber \\
\eta_2 &=& 3,\nonumber \\
\eta_3 &=& -3,
\label{eq:EVQ2}
\eea
\end{enumerate}
This point corresponds to a DM dominated Universe. Since the eigenvalues have mixed signs, the point is a saddle. Both points are shown in Table \ref{tab:sub_B}.
\begin{table}[tbp]
\centering
\begin{tabular}{|c|c|c|c|c|c|c|c|c|}
\hline
Point & \qquad $x$ & \qquad $y$ & \qquad $z$ & \qquad Stability & \qquad 
$w_{eff}$ \\
\hline
$Q_1$ & \qquad 0 & \qquad 0 & \qquad 0 &  \qquad Undetermined  & \qquad 
$-1$ \\
$Q_2$ & \qquad 0 &\qquad 1 &\qquad 0 &\qquad  Saddle  &\qquad 
0 \\
\hline
\end{tabular}
\label{tab:sub_B}
\caption{Critical points and cosmological parameter for Model 2, see subsection \ref{sec:sub_B}.}
\end{table}
\subsection{Model 3: $Q=\alpha \dot{\rho}_m$ with Dust Matter}
\label{sec:sub_C}
\begin{figure}[tbp]
\begin{center}
\begin{tabular}{ccc}
{\includegraphics[width=1.9in,height=2in,angle=0]{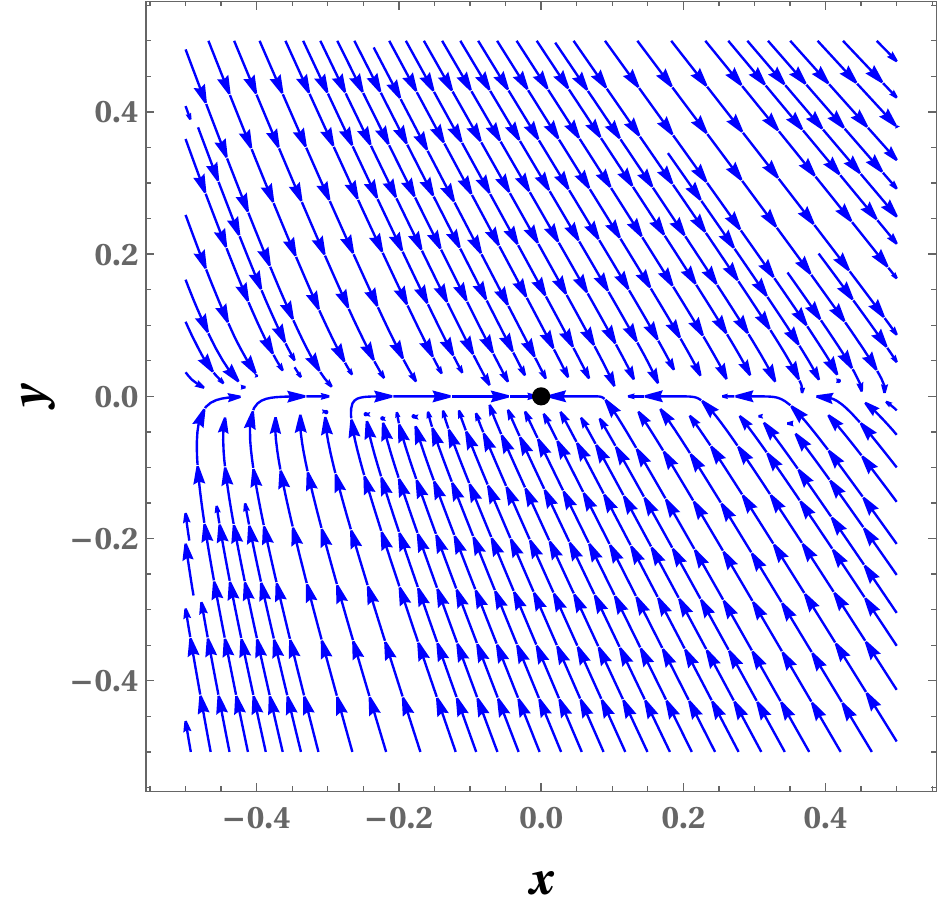}} &
{\includegraphics[width=1.9in,height=2in,angle=0]{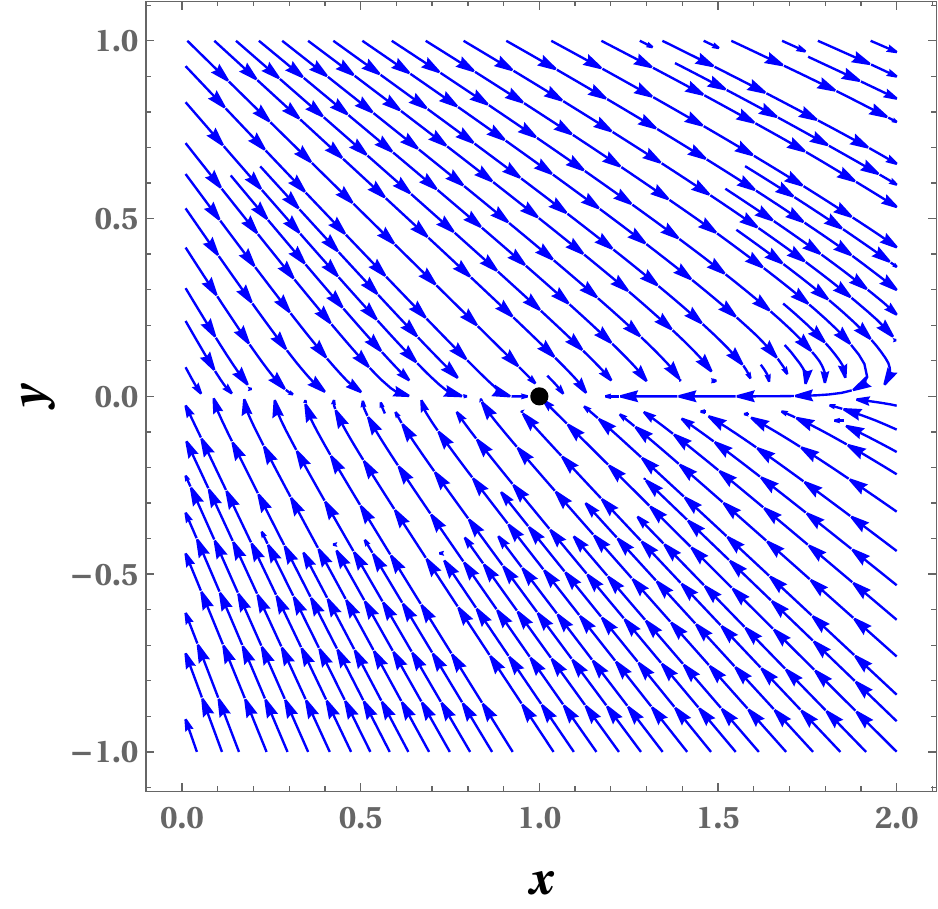}} &
{\includegraphics[width=1.9in,height=2in,angle=0]{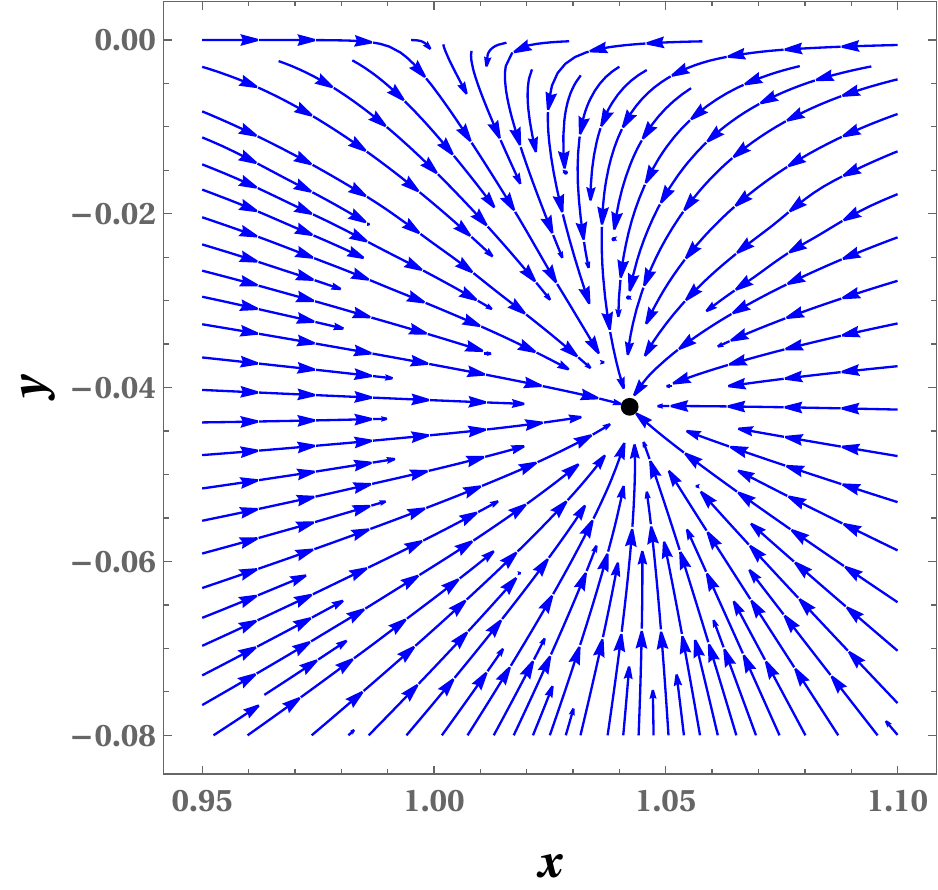}} 
\end{tabular}
\end{center}
\caption{The figure depicts the stable fixed points corresponding to $R_1$ (left panel; $\alpha=0.5$, $w_x=-0.95$ ), $R_2$ (middle panel; $\alpha=0.5$, $w_x=-1.1$), and $R_3$ (right panel; $\alpha=20$, $w_x=-1.01$ ) for Model 3 of subsection \ref{sec:sub_C}.  Under the chosen parameters, the associated eigenvalues are negative, indicating that the fixed point is a stable attractor, specifically an attractive node. The black dots mark the stable attractors in the phase space. } 
\label{fig:R}
\end{figure}

In this subsection, we consider the interaction term $Q=\alpha \dot{\rho}_m$ in the presence of pressureless matter ($p_m=0$), commonly referred to as dust matter in the literature. For this choice of matter component, the autonomous system (\ref{eq:auto}) becomes independent of the variable $z$, thereby reducing the dynamical system to two dimensions:
\begin{align}
\frac{dx}{dN} &= -3(1+w_x)x-\frac{\kappa^2 Q}{3 H^3}-2x \frac{\dot{H}}{H^2},\nonumber\\
\frac{dy}{dN} &= -3y+\frac{\kappa^2 Q}{3 H^3}-2y \frac{\dot{H}}{H^2},
\label{eq:auto2}
\end{align}
where,
\begin{eqnarray}
\label{eq:wd}
w_x &=& \frac{p_d}{\rho_d},\\
\label{eq:KQ1_DM}
\frac{\kappa^2Q}{3H^3} &=& \frac{3\alpha y}{\alpha-1},\\
\frac{\dot{H}}{H^2} &=& -\frac{3}{2}\Big{(}(1+w_x)x+y \Big{)}
\label{eq:Hd1_DM}
\end{eqnarray}
The stationary points are obtained by setting the left-hand side of the autonomous system (\ref{eq:auto2}) equal to zero. The resulting critical points and their stability properties are summarized below.
\begin{enumerate}
\item[\textbf{$R_1$:}] 
    \bea
   x = 0,  \qquad y = 0, 
    \eea
The eigenvalues of the Jacobian matrix at this point are
\bea
\eta_1 &=& - 3 (1+w_x), \nonumber \\
\eta_2 &=& \frac{3}{\alpha-1}, 
\label{eq:EVR1}
\eea
This point represents a vacuum-like state. For $\alpha<1$ and $w_x>-1$, both eigenvalues are negative, making $R_1$ a stable attractor.
\item[\textbf{$R_2$:}]
    \bea
    x=1, \qquad y =0, 
    \eea
The corresponding eigenvalues are
    \bea
\eta_1 &=& 3 (1+w_x),\nonumber \\
\eta_2 &=& \frac{3(w_x \alpha -w_x+ \alpha)}{\alpha-1}
\label{eq:EVR2}
\eea
This is a DE dominated solution. It becomes a stable node for $\alpha<1$ and $w_x<-1$.
    \item[\textbf{$R_3$:}] 
    \bea
   x=\frac{\alpha}{w_x(1-\alpha)}, \qquad y=\frac{w_x- w_x \alpha - \alpha}{w_x(1-\alpha)}, 
    \eea
The eigenvalues associated with this critical point are
\bea
\eta_1 &=& \frac{3}{1-\alpha},\nonumber \\
\eta_2 &=& \frac{3(w_x- w_x \alpha - \alpha)}{\alpha-1}
\label{eq:EVR3}
\eea
This scaling solution contains nonzero contributions from both DM and DE. Under $\alpha>1$ and $w_x>\frac{\alpha}{1-\alpha}$, it acts as a stable attractor.
\item[\textbf{$R_4$:}]
    \bea
    x=0, \qquad y =1, 
    \eea
The corresponding eigenvalues are
    \bea
\eta_1 &=& 3,\nonumber \\
\eta_2 &=& \frac{3(w_x  -w_x \alpha + \alpha)}{\alpha-1}
\label{eq:EVR4}
\eea
\end{enumerate}
This point corresponds to a matter dominated Universe. Since one eigenvalue is always positive, it is a saddle point. The above analysis shows that the dynamical system admits three possible stable nodes, namely $R_1$, $R_2$ and $R_3$, in different regions of the parameter space. In contrast, $R_4$ is always a saddle point due to the presence of a positive eigenvalue. These results indicate that the late-time behavior of the system strongly depends on the interaction parameter $\alpha$ and the DE equation of state parameter $w_x$. These points are displayed in Table \ref{tab:sub_C}. The phase portraits for points $R_1$, $R_2$ and $R_3$ are depicted in Fig. \ref{fig:R}.
\begin{table}[tbp]
\centering
\begin{tabular}{|c|c|c|c|c|c|c|c|c|}
\hline
Point & \qquad $x$ & \qquad $y$  & \qquad Stability & \qquad 
$w_{eff}$ \\
\hline
$R_1$ & \qquad 0 & \qquad 0  &\qquad  Stable  &\qquad 
 $-1$ \\
$R_2$ & \qquad 1 &\qquad 0  &\qquad  Stable  &\qquad 
$w_x$ \\
$R_3$ &\qquad  $\frac{\alpha}{w_x(1-\alpha)}$ &\qquad $\frac{w_x- w_x \alpha - \alpha}{w_x(1-\alpha)}$ &\qquad  Stable &\qquad 
 $\frac{\alpha}{1-\alpha}$ \\
$R_4$ &\qquad 0 &\qquad 1  &\qquad  Saddle  &\qquad  
 0 \\
\hline
\end{tabular}
\label{tab:sub_C}
\caption{Critical points and stability for Model 3, see subsection \ref{sec:sub_C}.}
\end{table}
\subsection{Model 4: $Q=\beta \dot{\rho}_d$ with Dust Matter}
\label{sec:sub_D}
\begin{figure}[tbp]
\begin{center}
\begin{tabular}{ccc}
{\includegraphics[width=1.9in,height=2in,angle=0]{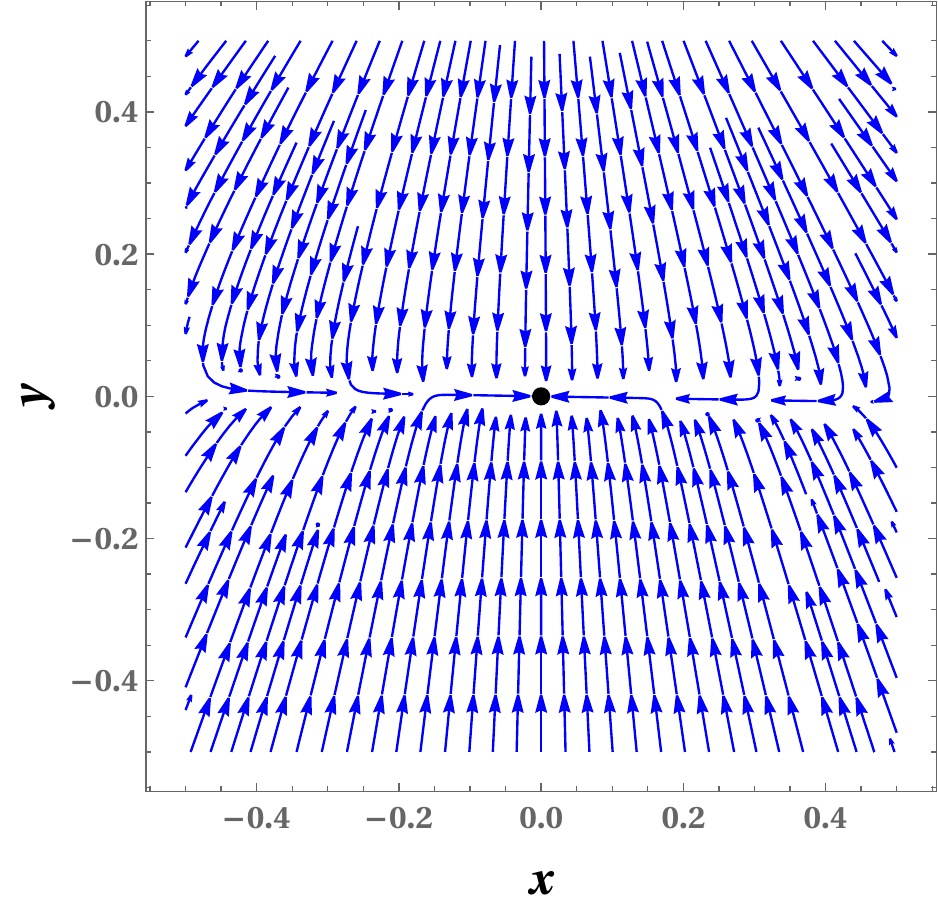}} &
{\includegraphics[width=1.9in,height=2in,angle=0]{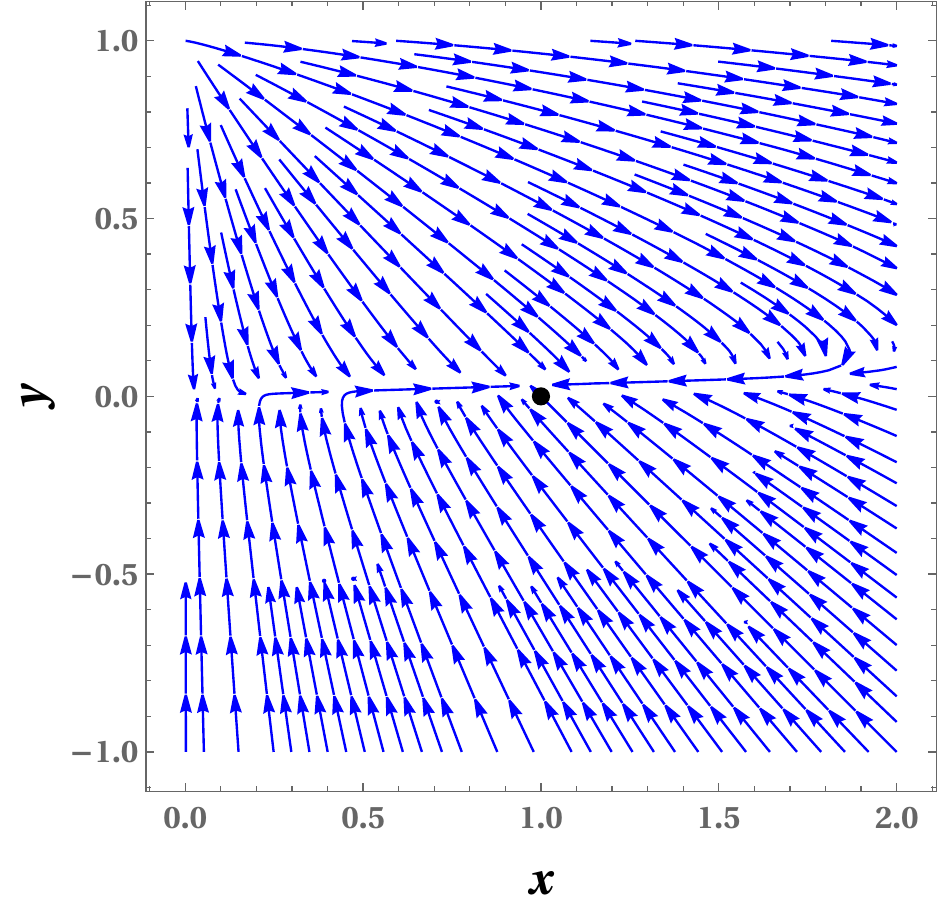}} &
{\includegraphics[width=1.9in,height=2in,angle=0]{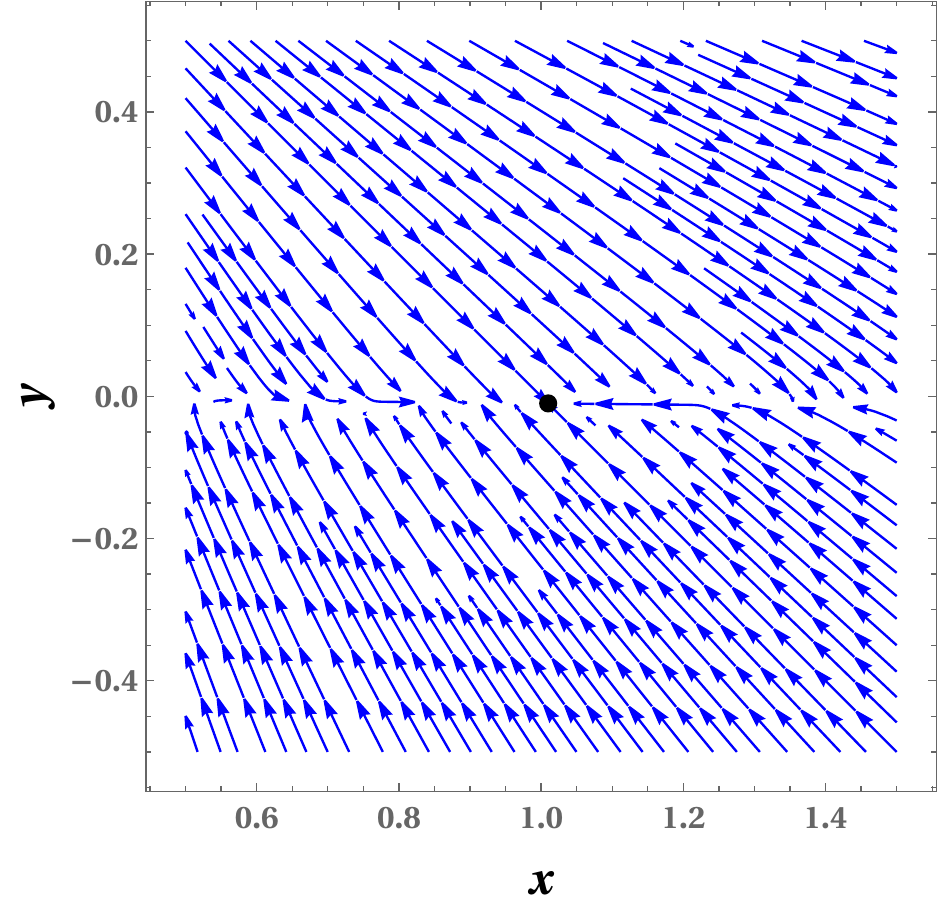}} 
\end{tabular}
\end{center}
\caption{The figure exhibits the stable fixed points corresponding to $S_1$ (left panel; $\beta=0.5$, $w_x=-0.95$), $S_2$ (middle panel; $\beta=0.5$, $w_x=-1.1$), and $S_3$ (right panel; $\beta=-0.5$, $w_x=-1.01$) for Model 4 of subsection \ref{sec:sub_D}.  For given parameters, the corresponding eigenvalues are negative, indicating that the fixed point is an attractive node. The black dots represent the stable attractors, where all trajectories of phase space converge. } 
\label{fig:S}
\end{figure}

Following the analysis presented in Subsection \ref{sec:sub_C}, we consider the interaction term $Q=\beta \dot{\rho}_d$ with $p_m=0$. The autonomous system given by Eq. (\ref{eq:auto2}) and Eq. (\ref{eq:Hd1_DM}) remain unchanged. However, owing to the modified coupling function, Eq. (\ref{eq:KQ1_DM}) is altered and takes the form
\begin{eqnarray}
\frac{\kappa^2Q}{3H^3} &=& -\frac{3 \beta (1+w_x) x}{1+\beta}
\label{eq:KQ2_DM}
\end{eqnarray}
To identify the stationary points and investigate their stability characteristics, we analyze the autonomous system (\ref{eq:auto2}) in conjunction with Eqs. (\ref{eq:Hd1_DM}) and (\ref{eq:KQ2_DM}). The resulting critical points are listed below.
\begin{enumerate}
\item[\textbf{$S_1$:}] 
    \bea
    x=0, \qquad y =0,
    \eea
The corresponding eigenvalues are
    \bea
\eta_1 &=& -3,\nonumber \\
\eta_2 &=& -\frac{3(1+ w_x)}{1+\beta}
\label{eq:EVS1}
\eea
This vacuum-like solution is stable for $\beta>-1$ and $w_x>-1$.
\item[\textbf{$S_2$:}]
    \bea
    x=1, \qquad y =0,
    \eea
The corresponding eigenvalues are
    \bea
\eta_1 &=& 3(1+w_x),\nonumber \\
\eta_2 &=& \frac{3(w_x+2w_x \beta + \beta)}{1+\beta}
\label{eq:EVS2}
\eea
This DE dominated state becomes stable when $\beta\geq -1/2$ and $w_x<-1$.
    \item[\textbf{$S_3$:}] 
    \bea
   x=\frac{w_x - \beta}{w_x(1+ \beta)}, \qquad y=\frac{(1+w_x) \beta}{w_x(1+ \beta)}, 
    \eea
The eigenvalues corresponding to this critical point are
\bea
\eta_1 &=& \frac{3(1+w_x)}{1+ \beta},\nonumber \\
\eta_2 &=&\frac{3(w_x-\beta)}{1+ \beta},
\label{eq:EVS3}
\eea
This scaling solution contains both DE and DM components. For $\beta>-1$ and $w_x<-1$,  it is stable and can describe the late-time accelerated Universe.
\item[\textbf{$S_4$:}] 
    \bea
   x=0, \qquad y=1, 
    \eea
\end{enumerate}
The eigenvalues corresponding to this critical point are
\bea
\eta_1 &=& 3,\nonumber \\
\eta_2 &=&-\frac{3(w_x-\beta)}{1+ \beta},
\label{eq:EVS4}
\eea
This matter dominated solution is always a saddle point because one eigenvalue remains positive. These points are summarized in Table \ref{tab:sub_D}. The phase space trajectories for points $S_1$, $S_2$ and $S_3$ are presented in Fig. \ref{fig:S}.
\begin{table}[tbp]
\centering
\begin{tabular}{|c|c|c|c|c|c|c|c|c|}
\hline
Point & \qquad $x$ & \qquad $y$ &  \qquad Stability & \qquad 
$w_{eff}$ \\
\hline
$S_1$ & \qquad 0 &\qquad 0 &\qquad   Stable  &\qquad 
$-1$ \\
$S_2$ &\qquad 1 &\qquad 0 &\qquad   Stable  &\qquad 
$w_x$ \\
$S_3$ &\qquad  $\frac{w_x - \beta}{w_x(1+ \beta)}$ &\qquad $\frac{(1+w_x) \beta}{w_x(1+ \beta)}$ & \qquad Stable  &\qquad 
$\frac{w_x - \beta}{1+ \beta}$ \\
$S_4$ &\qquad 0 &\qquad 1 &\qquad   Saddle  &\qquad 
 0 \\
\hline
\end{tabular}
\label{tab:sub_D}
\caption{Stationary points and cosmological parameter for Model 4 of subsection \ref{sec:sub_D}.}
\end{table}

The phase-space analysis reveals a clear distinction between the SFDM and dust matter scenarios. For the SFDM (Models 1 and 2), the dynamical system admits only non-hyperbolic or saddle critical points, preventing a definitive identification of late-time attractors through linear stability analysis. In contrast, the dust-matter models (Models 3 and 4) possess stable critical points for appropriate ranges of the interaction parameters and DE equation of state parameter, indicating the existence of viable late-time attractor solutions. These results demonstrate that the nature of the dark matter component and the form of the interaction term significantly influence the asymptotic cosmological dynamics.
\section{Loop Quantum Cosmology Framework}
\label{sec:LQC}
We perform a comprehensive dynamical analysis of the previous four models within the framework of LQC, and emphasis on the existence and stability of late-time attractor solutions. Furthermore, we investigate whether the singular behaviors encountered in the classical dynamical system are alleviated or persist in the quantum-corrected scenario. Before proceeding with the LQC analysis, it is worthwhile to briefly review the motivation for considering SFDM. Originally proposed in Ref. \cite{Berezhiani:2015bqa} and subsequently developed in Refs. \cite{Berezhiani:2015pia,Hodson:2016rck,Berezhiani:2017tth}, the SFDM paradigm provides an intriguing framework that successfully reproduces $\Lambda$CDM cosmology on large scales while simultaneously accounting for the phenomenological successes of Modified Newtonian Dynamics (MOND) at galactic scales. In this context, DM consists of strongly self-interacting axion-like particles with masses of the order of an electron volt. Under appropriate conditions, these particles condense into a superfluid phase whose effects become significant on galactic scales, while remaining negligible at the scale of galaxy clusters.

We briefly summarize the essential aspects of LQC. For comprehensive reviews and detailed discussions, see Refs. \cite{LQC1,LQC3,LQC4,LQC5,Salo:2016dsr,Xiong:2007cn,Amoros:2014tha,Cai:2014zga,deHaro:2014kxa,Kleidis:2018plu,c1,c2,c3,c4,c5,Kleidis:2017ftt}. Throughout this work, we focus on holonomy-corrected LQC, in which the underlying spacetime geometry is discrete and quantum gravitational effects are incorporated through holonomy corrections in the Hamiltonian formulation. In a spatially flat, homogeneous, and isotropic FLRW universe, the effective Hamiltonian in LQC is given by
\begin{equation}
\label{2.10a}
\mathcal{H}=-\frac{3v\sin^{2}(\lambda b)}{8\pi G \gamma^{2}\lambda^{2}}
+\mathcal{H}_{M},
\end{equation}
where $\mathcal{H}_{M}$ denotes the matter Hamiltonian, $v=v_{0}a^{3}$, $v_{0}$ is the volume of the fiducial cell, $`a'$ represents scale factor, and $\kappa^2=8 \pi G$. The Barbero--Immirzi parameter $\gamma$, whose value is fixed from black-hole thermodynamics in Loop Quantum Gravity, is approximately $\gamma \simeq 0.2375$ \cite{75,76}.
The Hamiltonian equations governing the dynamical variables $v$ and $b$ are
\begin{eqnarray}
\label{LQCa}
\dot v &=& \frac{3v}{2\lambda\gamma}\sin(2\lambda b), \\
\label{LQCb}
\dot b &=& -\frac{3\sin^{2}(\lambda b)}{2\gamma\lambda^{2}}
-4\pi G\gamma P,
\end{eqnarray}
where $P$ denotes the total pressure of the cosmic fluid. The energy density is obtained from the Hamiltonian constraint $\mathcal{H}=0$, yielding
\begin{equation}
\label{2.11}
\rho=\rho_{c}\sin^{2}(\lambda b),
\end{equation}
where $\rho_{c}=\frac{3}{8\pi G\lambda^{2}\gamma^{2}}$ is the critical energy density. Unlike standard cosmology, where the Big Bang is associated with an initial singularity, LQC incorporates non-perturbative quantum geometric effects that modify the classical gravitational dynamics. As a consequence, the Einstein field equations are replaced by effective cosmological equations that remain well defined even in the high-energy regime. Using Eqs. (\ref{LQCa}) and (\ref{LQCb}), one obtains the modified Friedmann and Raychaudhuri equations:
\begin{eqnarray}
H^{2} &=& \frac{\kappa^2}{3}\rho \left(1-\frac{\rho}{\rho_{c}}\right),
\label{eq:Htot_LQC}
\end{eqnarray}
Here, the Hubble parameter is defined as
$ H\equiv \frac{\dot a}{a} =\frac{\dot v}{3v}$. The $\rho$ denotes the total energy density of the cosmic fluid, $\rho=\rho_d+\rho_m$. The correction factor $\left(1-\rho/\rho_{c}\right)$ encapsulates the leading quantum geometric effects, and becomes significant when the energy density approaches the critical value $\rho_{c}$. At low densities $(\rho \ll \rho_{c})$, the classical Friedmann equation (\ref{eq:H-CC}) of GR is recovered. However, as $\rho \rightarrow \rho_{c}$, quantum effects generate a repulsive gravitational component that drives the Hubble parameter to zero, thereby halting the contraction of the universe and replacing the classical Big Bang singularity with a non-singular quantum bounce. Consequently, the effective LQC framework provides a consistent description of the early universe while smoothly connecting quantum gravitational dynamics with the classical cosmological evolution at late times. We now extend the dynamical analysis to the framework of LQC and derive the corresponding autonomous system for the IDE in a spatially flat FLRW universe. The continuity equations for the DE and SFDM components remain unchanged and are given by Eqs. (\ref{eq:conser_1}) and (\ref{eq:conser_2}). Likewise, the interaction term $Q$ is assumed to retain the general form. Differentiating Eq. (\ref{eq:Htot_LQC}) with respect to cosmic time and employing the continuity equations yields
\begin{equation}
\label{eq:Htot_d_LQC}
\dot{H}=-\frac{\kappa^2}{2} \left(\rho_m+\rho_d+p_m +p_d\right)
\left(1-2\frac{\rho_m+\rho_d}{\rho_c}\right),
\end{equation}
In the following subsections, we extend the dynamical analysis of the interacting dark-sector, and consider the same four interaction scenarios examined in the classical case, namely $Q=\alpha\dot{\rho}_m$ and $Q=\beta\dot{\rho}_d$, with SFDM and PM.
\subsection{Model 1: $Q=\alpha \dot{\rho}_m$ with  Superfluid Dark Matter}
\label{sec:sub_A_LQC}
For SFDM models, the DM and DE Eqs. (\ref{eq:pm_GR}) and (\ref{eq:pd_GR}) have the following forms \cite{77}.
\begin{eqnarray}
\label{eq:pm_LQC}
p_m &=& \frac{B}{\rho_c^2}\rho_m^3,\\
\label{eq:pd_LQC}
p_d &=& -\rho_d -w_d\rho_d -\frac{A}{\rho_c}\rho_d^2,
\end{eqnarray}
where $A$, $B$, and $w_d$ are free model parameters. We first consider the interaction $Q=\alpha\dot{\rho}_m$ in the SFDM scenario. Our objective is to formulate an autonomous system analogous to that presented in Eq. (\ref{eq:auto}). To this end, we introduce the  dimensionless variables:
\begin{equation}
\label{eq:variable_LQC}
x=\frac{\kappa^2\rho_d}{3H^2},
\qquad
y=\frac{\kappa^2\rho_m}{3H^2},
\qquad
z=\frac{H^2}{\kappa^2\rho_c}.
\end{equation}
The variables $x$ and $y$ represent the normalized energy densities of DE and DM, respectively, while the variable $z$ quantifies the magnitude of quantum gravitational corrections. These variables provide a convenient framework for recasting the cosmological evolution equations into an autonomous system.
\begin{align}
\frac{dx}{dN} &= 3 w_d x +9Ax^2z-\frac{\kappa^2 Q}{3 H^3}-2x \frac{\dot{H}}{H^2},\nonumber\\
\frac{dy}{dN} &= -3y-27By^3 z^2+\frac{\kappa^2 Q}{3 H^3}-2y \frac{\dot{H}}{H^2},\nonumber\\
\frac{dz}{dN} &= 2z \frac{\dot{H}}{H^2},
\label{eq:auto_LQC}
\end{align}
where, $\frac{\kappa^2Q}{3H^3}$ is given by Eq. (\ref{eq:KQ1_SDF}), and 
\begin{eqnarray}
\frac{\dot{H}}{H^2} &=& -\frac{3}{2}\Big{(}y-w_d x -3Ax^2 z \Big{)} \Big{(}1-6 z(x+y) \Big{)}
\label{eq:Hd1_LQC}
\end{eqnarray}
We determine the critical points by imposing the stationary conditions of the autonomous system and study their stability through the eigenvalues of the corresponding Jacobian matrix. The resulting phase-space structure allows us to identify possible attractor solutions and assess the impact of quantum corrections on the late-time dynamics.
\begin{enumerate}
\item[\textbf{$T_1$:}] 
    \bea
   x = 0,  \qquad y = 0, \qquad z = 0,
    \eea
The eigenvalues of the Jacobian matrix evaluated at this point are
\bea
\eta_1 &=& 0, \nonumber \\
\eta_2 &=& 3 w_d, \nonumber \\
\eta_3 &=& \frac{3}{\alpha-1}
\label{eq:EVT1}
\eea
This point corresponds to a vacuum-like configuration in which both DE and DM densities vanish. Since one eigenvalue is zero, the point is non-hyperbolic and its stability cannot be fully determined through linear analysis.
\item[\textbf{$T_2$:}]
    \bea
    x=1, \qquad y =0 , \qquad  z = 0, 
    \eea
The corresponding eigenvalues are
    \bea
\eta_1 &=& 3 w_d,\nonumber \\
\eta_2 &=& -3 w_d,\nonumber \\
\eta_3 &=& \frac{3(1+w_d - w_d \alpha)}{\alpha-1}
\label{eq:EVT2}
\eea
This is a DE dominated solution with $w_{eff}=-1-w_d$. For $\alpha>1$ and $w_d > \frac{1}{\alpha-1}$, the coexistence of positive and negative eigenvalues indicates both attracting and repelling directions in phase space; therefore, $T_2$ is a saddle point.
\item[\textbf{$T_3$:}]
    \bea
    x=\frac{\alpha}{(1+w_d)(\alpha-1)}, \qquad y =-\frac{1+w_d - w_d \alpha}{(1+w_d)(\alpha-1)} , \qquad  z = 0, 
    \eea
The corresponding eigenvalues are
    \bea
\eta_1 &=& \frac{3}{\alpha-1},\nonumber \\
\eta_2 &=& -\frac{3}{\alpha-1},\nonumber \\
\eta_3 &=& -\frac{3(1+w_d - w_d \alpha)}{\alpha-1}
\label{eq:EVT3}
\eea
This scaling solution contains contributions from both dark sectors. The mixed signs of the eigenvalues for $\alpha>1$ and $w_d > \frac{1}{\alpha-1}$, imply that it is a saddle point.
  \item[\textbf{$T_4$:}] 
   \bea
  x=0, \qquad y=1, \qquad z = 0, 
   \eea
The eigenvalues associated with this critical point are
\bea
\eta_1 &=& 3,\nonumber \\
\eta_2 &=& - 3,\nonumber \\
\eta_3 &=& 3(1+w_d),
\label{eq:EVT4}
\eea
This point represents a SFDM dominated Universe. Since the eigenvalues have mixed signs, the solution is unstable and behaves as a saddle point. All the points are summarized in Table \ref{tab:sub_A_LQC}. In this subsection, we do not find any stable point for SFDM.
\end{enumerate}
\begin{table}[tbp]
\centering
\begin{tabular}{|c|c|c|c|c|c|c|c|c|}
\hline
Point & \qquad $x$ & \qquad $y$ & \qquad $z$ & \qquad Stability & \qquad 
$w_{eff}$ \\
\hline
$T_1$ &\qquad  0 &\qquad 0 &\qquad 0 &\qquad  Undetermined   &\qquad $-1$ \\
$T_2$ &\qquad  1 &\qquad 0 &\qquad 0 &\qquad  Saddle  &\qquad $-1-w_d$ \\
$T_3$ &\qquad  $\frac{\alpha}{(1+w_d)(\alpha-1)}$ &\qquad $-\frac{1+w_d- w_d \alpha}{(1+w_d)(\alpha-1)}$ &\qquad 0 &\qquad  Saddle  &\qquad $\frac{\alpha}{1-\alpha}$ \\
$T_4$ &\qquad 0 &\qquad 1 &\qquad 0 &\qquad  Saddle  &\qquad 0 \\
\hline
\end{tabular}
\label{tab:sub_A_LQC}
\caption{Critical points and cosmological parameter for subsection \ref{sec:sub_A_LQC}.}
\end{table}
\subsection{Model 2: $Q=\beta \dot{\rho}_d$ with  Superfluid Dark Matter}
\label{sec:sub_B_LQC}
We now replace the interaction term by $Q=\beta\dot{\rho}_d$. The autonomous system (\ref{eq:auto_LQC}) and Eq. (\ref{eq:Hd1_LQC}) remain unchanged except for the interaction contribution, which is given by 
\begin{eqnarray}
\frac{\kappa^2Q}{3H^3} &=& \frac{3 \beta( w_d x + 3 A z x^2)}{1+\beta}
\label{eq:KQ2_SDF_LQC}
\end{eqnarray}
To determine the stationary points and examine their stability properties, we analyze Eq. (\ref{eq:auto_LQC}) together with Eqs. (\ref{eq:Hd1_LQC}) and (\ref{eq:KQ2_SDF_LQC}). The corresponding critical points are listed below.
\begin{enumerate}
\item[\textbf{$U_1$:}] 
    \bea
   x = 0,  \qquad y = 0, \qquad z = 0,
    \eea
The eigenvalues of the Jacobian matrix evaluated at this point are
\bea
\eta_1 &=& 0, \nonumber \\
\eta_2 &=& -3, \nonumber \\
\eta_3 &=& \frac{3 w_d}{1+\beta}
\label{eq:EVU1}
\eea
This vacuum solution is non-hyperbolic due to the presence of a vanishing eigenvalue. Its stability requires methods beyond linear stability theory.
\item[\textbf{$U_2$:}]
    \bea
    x=1, \qquad y =0 , \qquad  z = 0, 
    \eea
The corresponding eigenvalues are
    \bea
\eta_1 &=& 3 w_d,\nonumber \\
\eta_2 &=& -3 w_d,\nonumber \\
\eta_3 &=& -\frac{3(1+\beta + w_d + 2 w_d \beta)}{1+\beta}
\label{eq:EVU2}
\eea
This DE dominated state possesses both stable and unstable directions for $\beta>-\frac{1}{2}$ and $w_d>-\frac{1+\beta}{1+2 \beta}$, and is therefore a saddle point.
    \item[\textbf{$U_3$:}] 
    \bea
   x=\frac{1+\beta + w_d}{(1+w_d)(1+ \beta)}, \qquad y=\frac{w_d \beta}{(1+w_d)(1+ \beta)}, \qquad z = 0, 
    \eea
The eigenvalues corresponding to this critical point are
\bea
\eta_1 &=& \frac{3 w_d}{1+ \beta},\nonumber \\
\eta_2 &=& - \frac{3 w_d}{1+ \beta},\nonumber \\
\eta_3 &=&-\frac{3(1+ w_d +\beta)}{1+ \beta},
\label{eq:EVU3}
\eea
For $\beta>-1$ and $w_d>-1-\beta$, this scaling solution contains nonzero DE and DM components. The eigenvalues exhibit mixed signs, indicating saddle behavior. 
\item[\textbf{$U_4$:}]
    \bea
    x=0, \qquad y =1, \qquad  z = 0, 
    \eea
The corresponding eigenvalues are
    \bea
\eta_1 &=& 3,\nonumber \\
\eta_2 &=& -3,\nonumber \\
\eta_3 &=& \frac{3(1+ w_d + \beta)}{1+\beta}
\label{eq:EVU2}
\eea
This matter dominated configuration is also a saddle point due to positive eigenvalue. These points are presented in Table \ref{tab:sub_B_LQC}. Similar to subsection \ref{sec:sub_A_LQC}, here also, we do not obtain any stable point for SFDM.
\end{enumerate}
\begin{table}[tbp]
\centering
\begin{tabular}{|c|c|c|c|c|c|c|c|c|}
\hline
Point & \qquad $x$ & \qquad $y$ & \qquad $z$ & \qquad Stability & \qquad 
$w_{eff}$ \\
\hline
$U_1$ & \qquad 0 &\qquad 0 &\qquad 0 &\qquad  Undetermined  &\qquad $-1$ \\
$U_2$ &\qquad 1 &\qquad 0 &\qquad 0 &\qquad  Saddle  &\qquad $-1-w_d$ \\
$U_3$ &\qquad  $\frac{1+\beta + w_d}{(1+w_d)(1+ \beta)}$ &\qquad $\frac{w_d \beta}{(1+w_d)(1+ \beta)}$ &\qquad 0 &\qquad  Saddle  &\qquad $-\frac{\beta+w_d+1}{1+\beta}$ \\
$U_4$ &\qquad 0 &\qquad 1 &\qquad 0 &\qquad  Saddle  &\qquad 0 \\
\hline
\end{tabular}
\label{tab:sub_B_LQC}
\caption{Critical points and cosmological parameter for subsection \ref{sec:sub_B_LQC}.}
\end{table}
\subsection{Model 3: $Q=\alpha \dot{\rho}_m$ with Dust Matter}
\label{sec:sub_C_LQC}
Next, we consider pressureless matter and interaction term $Q=\alpha\dot{\rho}_m$. The resulting autonomous system becomes
\begin{align}
\frac{dx}{dN} &= -3(1+w_x)x-\frac{\kappa^2 Q}{3 H^3}-2x \frac{\dot{H}}{H^2},\nonumber\\
\frac{dy}{dN} &= -3y+\frac{\kappa^2 Q}{3 H^3}-2y \frac{\dot{H}}{H^2},\nonumber\\
\frac{dz}{dN} &= 2z \frac{\dot{H}}{H^2},
\label{eq:auto2_LQC}
\end{align}
where, $\frac{\kappa^2Q}{3H^3}$ is given in Eq. (\ref{eq:KQ1_DM}), and
\begin{eqnarray}
\frac{\dot{H}}{H^2} &=& -\frac{3}{2}\Big{(}(1+w_x)x+y \Big{)} \Big{(}1-6z(x+y) \Big{)}
\label{eq:Hd1_DM_LQC}
\end{eqnarray}
The associated critical points and their stability characteristics are presented below.
\begin{enumerate}
\item[\textbf{$V_1$:}] 
    \bea
   x = 0,  \qquad y = 0,   \qquad z = 0, 
    \eea
The eigenvalues of the Jacobian matrix at this point are
\bea
\eta_1 &=& 0, \nonumber \\
\eta_2 &=& - 3 (1+w_x), \nonumber \\
\eta_3 &=& \frac{3}{\alpha-1}, 
\label{eq:EVV1}
\eea
This vacuum-like solution is non-hyperbolic owing to the presence of a zero eigenvalue. Consequently, its stability remains undetermined within linear theory.
\item[\textbf{$V_2$:}]
    \bea
    x=1, \qquad y =0,  \qquad z = 0, 
    \eea
The corresponding eigenvalues are
    \bea
\eta_1 &=& 3 (1+w_x),\nonumber \\
\eta_2 &=& -3 (1+w_x),\nonumber \\
\eta_3 &=& \frac{3(w_x \alpha -w_x+ \alpha)}{\alpha-1}
\label{eq:EVV2}
\eea
This DE dominated state has both stable and unstable directions and is therefore classified as a saddle point.
    \item[\textbf{$V_3$:}] 
    \bea
   x=\frac{\alpha}{w_x(1-\alpha)}, \qquad y=\frac{w_x- w_x \alpha - \alpha}{w_x(1-\alpha)}, \qquad z = 0, 
    \eea
The eigenvalues associated with this critical point are
\bea
\eta_1 &=& \frac{3}{1-\alpha},\nonumber \\
\eta_2 &=& -\frac{3}{1-\alpha},\nonumber \\
\eta_3 &=& \frac{3(w_x \alpha + \alpha - w_x )}{w_x(1-\alpha)}
\label{eq:EVV3}
\eea
This scaling solution describes a Universe in which DE and DM coexist. The mixed-sign eigenvalue spectrum renders it a saddle point.
\item[\textbf{$V_4$:}]
    \bea
    x=0, \qquad y =1, \qquad z = 0, 
    \eea
The corresponding eigenvalues are
    \bea
\eta_1 &=& 3,\nonumber \\
\eta_2 &=& -3,\nonumber \\
\eta_3 &=& \frac{3(w_x \alpha- w_x - \alpha)}{1-\alpha}
\label{eq:EVV4}
\eea
\end{enumerate}
This matter dominated solution contains an unstable direction in phase space and therefore behaves as a saddle point. All points are presented in Table \ref{tab:sub_C_LQC}. In this case, all the critical points are saddle for PM also.
\begin{table}[tbp]
\centering
\begin{tabular}{|c|c|c|c|c|c|c|c|c|}
\hline
Point & \qquad $x$ & \qquad $y$ & \qquad $z$ & \qquad Stability & \qquad 
$w_{eff}$ \\
\hline
$V_1$ &\qquad  0 &\qquad 0 &\qquad 0 &\qquad  Undetermined  &\qquad $-1$ \\
$V_2$ & \qquad 1 &\qquad 0 &\qquad 0 &\qquad  Saddle  &\qquad $w_x$  \\
$V_3$ &\qquad  $\frac{\alpha}{w_x(1-\alpha)}$ &\qquad $\frac{w_x- w_x \alpha - \alpha}{w_x(1-\alpha)}$ &\qquad 0 &\qquad  Saddle  &\qquad $\frac{\alpha}{1-\alpha}$ \\
$V_4$ &\qquad 0 &\qquad 1  &\qquad 0 &\qquad  Saddle  &\qquad 0  \\
\hline
\end{tabular}
\label{tab:sub_C_LQC}
\caption{Critical points and cosmological parameter for subsection \ref{sec:sub_C_LQC}.}
\end{table}
\subsection{Model 4: $Q=\beta \dot{\rho}_d$ with Dust Matter}
\label{sec:sub_D_LQC}
Finally, we analyze the dust matter with interaction term $Q=\beta\dot{\rho}_d$. Using Eqs. (\ref{eq:auto2_LQC}), (\ref{eq:Hd1_DM_LQC}), and (\ref{eq:KQ2_DM}), we obtain the critical points and determine their stability through the eigenvalues of the Jacobian matrix.
\begin{enumerate}
\item[\textbf{$W_1$:}] 
    \bea
    x=0, \qquad y =0, \qquad z =0,
    \eea
The corresponding eigenvalues are
    \bea
\eta_1 &=& 0,\nonumber \\
\eta_2 &=& -3,\nonumber \\
\eta_3 &=& -\frac{3(1+ w_x)}{1+\beta}
\label{eq:EVW1}
\eea
This point corresponds to a vacuum state and is non-hyperbolic due to the existence of a zero eigenvalue.
\item[\textbf{$W_2$:}]
    \bea
    x=1, \qquad y =0,  \qquad z =0,
    \eea
The corresponding eigenvalues are
    \bea
\eta_1 &=& 3(1+w_x),\nonumber \\
\eta_2 &=& -3(1+w_x),\nonumber \\
\eta_3 &=& \frac{3(w_x+2w_x \beta + \beta)}{1+\beta}
\label{eq:EVW2}
\eea
This DE dominated solution is saddle because at least one eigenvalue is positive.
    \item[\textbf{$W_3$:}] 
    \bea
   x=\frac{w_x - \beta}{w_x(1+ \beta)}, \qquad y=\frac{(1+w_x) \beta}{w_x(1+ \beta)}, \qquad z =0,
    \eea
The eigenvalues corresponding to this critical point are
\bea
\eta_1 &=& \frac{3(1+w_x)}{1+ \beta},\nonumber \\
\eta_2 &=& -\frac{3(1+w_x)}{1+ \beta},\nonumber \\
\eta_3 &=&\frac{3(w_x-\beta)}{1+ \beta},
\label{eq:EVW3}
\eea
This scaling solution contains both DE and DM components. The presence of positive and negative eigenvalues implies saddle behavior.
\item[\textbf{$W_4$:}] 
    \bea
   x=0, \qquad y=1, \qquad z =0,
    \eea
\end{enumerate}
The eigenvalues corresponding to this critical point are
\bea
\eta_1 &=& 3,\nonumber \\
\eta_2 &=& -3,\nonumber \\
\eta_3 &=&\frac{3(\beta- w_x)}{1+ \beta},
\label{eq:EVW4}
\eea
This matter dominated configuration always possesses an unstable direction and therefore represents a saddle point. We summarize these points in Table \ref{tab:sub_D_LQC}.
\begin{table}[tbp]
\centering
\begin{tabular}{|c|c|c|c|c|c|c|c|c|}
\hline
Point & \qquad $x$ & \qquad $y$ & \qquad $z$ & \qquad Stability & \qquad 
$w_{eff}$ \\
\hline
$S_1$ & \qquad 0 &\qquad 0 &\qquad 0 &\qquad   Undetermined  &\qquad $-1$\\
$S_2$ &\qquad 1 &\qquad 0 &\qquad 0 &\qquad   Saddle  &\qquad $w_x$\\
$S_3$ &\qquad  $\frac{w_x - \beta}{w_x(1+ \beta)}$ &\qquad $\frac{(1+w_x) \beta}{w_x(1+ \beta)}$ &\qquad 0 &\qquad  Saddle  &\qquad $\frac{w_x -\beta}{1+\beta}$\\
$S_4$ &\qquad 0 &\qquad 1 &\qquad 0 & \qquad  Saddle  & \qquad 0 \\
\hline
\end{tabular}
\label{tab:sub_D_LQC}
\caption{Critical points and cosmological parameter for subsection \ref{sec:sub_D_LQC}.}
\end{table}

Overall, unlike the corresponding classical dust matter models, the LQC corrected systems do not admit stable attractor solutions. Most critical points are saddle, while the vacuum solutions are inconclusive due to non-hyperbolic nature of points, and can not be determined through linear stability theory. This suggests that quantum-gravitational corrections significantly modify the asymptotic structure of the phase space and may prevent the emergence of conventional late-time attractors.

Finally, we make some comments on the constant and non-linear dark energy EoS. The use of a constant dark energy EoS was primarily motivated by the objective of obtaining analytically tractable autonomous systems and identifying the critical points and their stability properties. A constant EoS provides a useful baseline model against which the effects of the DM nature, dark-sector interaction, and LQC corrections can be isolated. Importantly, the constant EoS model is not intended to imply that the observationally preferred DE dynamics must be constant. Rather, it represents a reference case within the broader dynamical framework. The paper already considers, in addition to the constant EoS description, a generalized nonlinear DE equation of state. In the classical SFDM analysis this is given by $p_d=-\rho_d-A\kappa^4\rho_d^2,$ which introduces a density dependent EoS. We have therefore expanded the discussion to emphasize that the constant EoS case should be regarded as a phenomenological reference model rather than a complete description of an evolving DE sector. In the classical formulation,
\be
w\equiv\frac{p_d}{\rho_d}=-1-A\kappa^4\rho_d.
\ee
Therefore, unlike the constant EoS model, the $w$ evolves as the DE density evolves. The nonlinear term consequently provides a mechanism for dynamical DE behavior. The sign of \(A\) determines the side of the phantom divide on which this simple classical parametrization lies. For \(A>0\), \(w<-1\), whereas for \(A<0\), \(w>-1\), provided \(\rho_d>0\). Thus, for the particular classical nonlinear EoS adopted here with constant \(A\), a generic continuous crossing from \(w>-1\) to \(w<-1\) does not occur merely through the evolution of \(\rho_d\); instead, the model remains on one side of the divide for a fixed nonzero sign of \(A\).

The LQC formulation is somewhat more general. In this framewoek, the DE pressure is written as $p_d=-\rho_d-w_d\rho_d-\frac{A}{\rho_c}\rho_d^2,$
so that
\be
w=-1-w_d-\frac{A}{\rho_c}\rho_d.
\ee
In this case the EoS is explicitly density dependent, and the condition for a phantom divide crossing is
\be
w=-1 \quad \Longleftrightarrow \quad w_d+\frac{A}{\rho_c}\rho_d=0.
\ee
Consequently, if the evolution of \(\rho_d\) causes this condition to be satisfied, a crossing of \(w=-1\) can in principle occur, see Refs. \cite{m1,m2,m3,m4} for detailed discussions on phantom crossing. We emphasize that the fixed point analysis presented in the current manuscript was designed primarily to determine the critical points and their stability. It does not by itself establish a  phantom to quintessence transition. Moreover, the nonlinear EoS is density dependent, and there is no explicit \(w(z)\) trajectory or numerical crossing analysis. Therefore, we acknowledge this limitation, and do not claim such a transition as one of the established results of the present work.
\section{Cosmological Implications}
\label{sec:CI}
The present dynamical system analysis provides a comprehensive comparison between IDE and DM in the frameworks of classical gravity and LQC. The results demonstrate that both the nature of DM component and the underlying gravitational theory play decisive roles in determining the asymptotic evolution of the Universe. Within the classical framework, the dust matter models exhibit a rich phase-space structure containing stable late-time attractors, including DE dominated and scaling solutions. These attractors arise for suitable ranges of the interaction parameters, indicating that a non-gravitational interaction within the dark sector can naturally drive the Universe toward accelerated expansion. In particular, the scaling solutions are of considerable cosmological interest because they allow DM and DE to evolve with comparable energy densities over extended periods. Such behavior may alleviate the cosmological coincidence problem by reducing the need for fine-tuned initial conditions.

In contrast, when the DM component is described by the generalized superfluid equation of state, the classical systems admit only saddle or non-hyperbolic critical points. The absence of stable attractors suggests that the nonlinear pressure associated with SFDM significantly modifies the asymptotic cosmological evolution and prevents the emergence of conventional late-time epoch. This result indicates that either alternative interaction mechanisms or more general superfluid equations of state may be required to obtain observationally viable attractor solutions.

The incorporation of LQC substantially changes the global dynamical behavior. Quantum geometric corrections modify the Friedmann and Raychaudhuri equations through the characteristic correction factor $(1-\rho/\rho_c)$, leading to a nonsingular cosmological evolution in which the classical Big Bang singularity is replaced by a quantum bounce. Although these corrections successfully regularize the high-energy regime, they also alter the stability properties of the dynamical system. For all four interacting models, the classical stable attractors disappear, while most equilibrium points become saddle and the remaining vacuum configurations are non-hyperbolic. Consequently, the quantum corrected dynamics do not admit conventional late-time attractor solutions. These findings suggest that quantum gravitational effects influence not only the very early Universe but also the qualitative structure of the cosmological phase space. The interplay between dark-sector interactions and quantum geometry produces a fundamentally different dynamical picture compared with classical cosmology. Therefore, IDE models that are phenomenologically successful in General Relativity may require significant modifications when quantum gravitational corrections are taken into account. Overall, the analysis emphasizes that both the interaction mechanism and the microscopic properties of DM are crucial ingredients in determining the cosmological evolution. The dynamical systems approach adopted here provides a useful framework for identifying viable asymptotic behaviors of IDE models.

Role of quantum corrections and dark-sector interaction: The nonsingular bounce obtained in LQC originates from the quantum geometrical correction \(1-\rho/\rho_c\) in the effective Friedmann equation and is therefore not generated by the DE-DM interaction. The interaction terms \(Q=\alpha\dot\rho_m\) and \(Q=\beta\dot\rho_d\)  modify the individual evolution of the dark-sector densities and consequently affect the trajectories through phase space. Thus, LQC determines the nonsingular nature of the high density regime, while the interaction determines how energy is exchanged between DE and DM and can substantially modify the subsequent cosmological dynamics and stability structure. This distinction is particularly evident from the comparison between the classical and LQC systems: the classical interacting dust models possess stable late-time attractors, whereas the corresponding LQC systems do not.
\section{Conclusion}
\label{sec:conc}
In this work, we have investigated the cosmological dynamics of IDE and DM within both classical  gravity and LQC using the dynamical systems approach. Two phenomenological interaction terms, $Q=\alpha\dot{\rho}_m$ and $Q=\beta\dot{\rho}_d$ were considered together with two different DM descriptions: a generalized SFDM and the standard pressureless fluid. The corresponding autonomous  systems were constructed, their critical points and stability properties were analyzed through linear perturbation theory. Within the classical framework, the SFDM models possess only saddle and non-hyperbolic points, indicating the absence of stable asymptotic cosmological states. In contrast, the dust matter models ($p_m=0$) admit stable attractors for appropriate choices of the interaction parameters. These attractor solutions provide viable descriptions of the late-time acceleration. The phase-space analysis reveals a marked difference between the cosmological dynamics of the SFDM and dust matter scenarios. For the SFDM cases (Models 1 and 2), the autonomous system admits only saddle and non-hyperbolic critical points, making it impossible to establish the existence of late-time attractors. By contrast, the dust matter models (Models 3 and 4) exhibit stable critical points for suitable ranges of the interaction parameters and the dark energy equation of state parameter, thereby providing viable late-time attractor solutions. These findings highlight that both the nature of the DM component and the choice of interaction term play a crucial role in determining the asymptotic evolution of the Universe. In particular, the critical points $R_1$, $R_2$, and $R_3$ of Model 3, together with $S_1$, $S_2$, and $S_3$ of Model 4, behave as stable nodes within their respective parameter ranges. The corresponding phase portraits (Figs. \ref{fig:R} and \ref{fig:S}) clearly demonstrated that the trajectories in the phase space converged toward these points, confirming their role as late-time attractors and reinforcing the stability of the cosmological solutions.

The analysis was subsequently extended to LQC by incorporating the effective quantum corrected Friedmann dynamics. Quantum geometric effects replace the classical cosmological singularity with a nonsingular bounce and significantly modify the global phase-space structure. Unlike the classical case, the LQC systems do not admit stable late-time attractors for either the superfluid or dust matter scenarios. Most equilibrium points become saddle, while the vacuum solutions remain non-hyperbolic, demonstrating that quantum gravitational corrections substantially alter the asymptotic behavior of interacting dark-sectors.

The present study highlights the combined influence of dark-sector interactions, the physical nature of DM, and quantum gravitational effects on the evolution of the Universe. While interacting dust models in classical gravity naturally lead to stable accelerated expansion, the inclusion of loop quantum corrections produces qualitatively different cosmological dynamics and suppresses the emergence of conventional late-time attractor solutions.


\acknowledgments
\noindent This research was funded by the Science Committee of the Ministry of Science and Higher Education of the Republic of Kazakhstan (Grant No. AP23489289).  MS acknowledges Integral University, Lucknow for financial support through Seed Money Grant 2024-2025 (Project Sanction No.: IUL/ICEIR/SMP/2024-04) and MCN: IU/R\&D/2026-MCN0004974. MS and PKD also thanks the Inter-University Centre for Astronomy and Astrophysics (IUCAA), Pune for the hospitality and facilities under the visiting associateship program where the work was completed. 

\paragraph{ Declaration of competing interest:} The authors declare that they have no known competing financial interests or personal relationships that could have appeared to influence the work reported in this paper.

\paragraph{ Data availability:} No data was used for the research described in the article.

\end{document}